\documentclass{article}
\usepackage{amssymb}
\usepackage{amsfonts}
\usepackage{amsmath}

\newtheorem{theorem}{Theorem}
\newtheorem{acknowledgement}[theorem]{Acknowledgement}

\input{tcilatex}

\begin{document}

\title{Complexified von Roos Hamiltonian's $\eta $-weak-pseudo-Hermiticity,
isospectrality and exact solvability}
\author{Omar Mustafa$^{1}$ and S.Habib Mazharimousavi$^{2}$ \\
%EndAName
Department of Physics, Eastern Mediterranean University, \\
G Magusa, North Cyprus, Mersin 10,Turkey\\
$^{1}$E-mail: omar.mustafa@emu.edu.tr\\
$^{2}$E-mail: habib.mazhari@emu.edu.tr}
\maketitle

\begin{abstract}
A complexified von Roos Hamiltonian is considered and a Hermitian
first-order intertwining differential operator is used to obtain the related
position dependent mass $\eta $-weak-pseudo-Hermitian Hamiltonians. Using a
Liouvillean-type change of variables, the $\eta $-weak-pseudo-Hermitian von
Roos Hamiltonians $H_{x}$ are mapped into the traditional Schr\"{o}dinger
Hamiltonian form $H_{q}$, where exact isospectral correspondence between $%
H_{x}$ and $H_{q}$ is obtained. Under a "\emph{user -friendly"} position
dependent mass settings, it is observed that for each exactly-solvable $\eta 
$-weak-pseudo-Hermitian \emph{reference}-Hamiltonian $H_{q}$ there is a set
of exactly-solvable $\eta $-weak-pseudo-Hermitian \emph{isospectral} \emph{%
target}-Hamiltonians $H_{x}$. A non-Hermitian $\mathcal{PT}$-symmetric Scarf
II and a non-Hermitian periodic-type $\mathcal{PT}$-symmetric Samsonov-Roy
potentials are used as \emph{reference} models and the corresponding $\eta $%
-weak-pseudo-Hermitian \emph{isospectral target}-Hamiltonians are obtained.

\end{abstract}

\section{Introduction}

Subjected to von Roos constraint $\alpha +\beta +\gamma =-1;$ $\alpha ,\beta
,\gamma \in 
%TCIMACRO{\U{211d} }%
%BeginExpansion
\mathbb{R}
%EndExpansion
,$ the von Roos position-dependent-mass (PDM) Hamiltonian [1-12] reads%
\begin{equation}
H=-\partial _{x}\left( \frac{1}{M\left( x\right) }\right) \partial _{x}+%
\tilde{V}\left( x\right) ,
\end{equation}%
with%
\begin{equation}
\tilde{V}\left( x\right) =\frac{1}{2}\left( 1+\beta \right) \frac{M^{\prime
\prime }\left( x\right) }{M\left( x\right) ^{2}}-\left[ \alpha \left( \alpha
+\beta +1\right) +\beta +1\right] \frac{M^{\prime }\left( x\right) ^{2}}{%
M\left( x\right) ^{3}}+V\left( x\right) ,
\end{equation}%
and primes denote derivatives. An obvious profile change of the potential $%
\tilde{V}\left( x\right) $ obtains as $\alpha ,\beta ,$ and $\gamma $
change, manifesting in effect an ordering ambiguity conflict in the process
of choosing a unique kinetic energy operator%
\begin{equation}
T=-\frac{1}{2}\left[ M\left( x\right) ^{\alpha }\partial _{x}M\left(
x\right) ^{\beta }\partial _{x}M\left( x\right) ^{\gamma }+M\left( x\right)
^{\gamma }\partial _{x}M\left( x\right) ^{\beta }\partial _{x}M\left(
x\right) ^{\alpha }\right]
\end{equation}%
Hence, $\alpha ,\beta ,$ and $\gamma $ are usually called the von Roos
ambiguity parameters. Yet, such PDM-quantum-particles (i.e., $M\left(
x\right) =m_{\circ }m\left( x\right) $) are used in the energy density
many-body problem, in the determination of the electronic properties of
semiconductors and quantum dots [1-5].

Regardless of the continuity requirements on the wave function at the
boundaries of abrupt herterojunctions between two crystals [6] and/or
Dutra's and Almeida's [7] reliability test, there exist several suggestions
for the kinetic energy operator in (3). We may mention the Gora's and
Williams' ($\beta =\gamma =0,$ $\alpha =-1$) [8], Ben Daniel's and Duke's ($%
\alpha =\gamma =0,$ $\beta =-1$) [9], Zhu's and Kroemer's ($\alpha =\gamma
=-1/2,$ $\beta =0$) [10], Li's and Kuhn's ($\beta =\gamma =-1/2,$ $\alpha =0$%
) [11], and the very recent Mustafa's and Mazharimousavi's ($\alpha =\gamma
=-1/4,$ $\beta =-1/2$) [3]. Nevertheless, in this work we shall deal with
these orderings irrespective to their classifications of being "good-"
(i.e., satisfying the continuity requirements on the wave function,
mentioned above, and surviving the Dutra's and Almeida's [7] reliability
test) or "to-be-discarded-" orderings (i.e., not satisfying the continuity
requirements on the wave function and/or failing the Dutra's and Almeida's
[7] reliability test). The reader is advised to refer to, e.g., Mustafa and
Mazharimousavi [3] for more details.

The growing interest in the non-Hermitian pseudo-Hermitian Hamiltonians with
real spectra [13-21], on the other hand, have inspired our resent work on
PDM first-order-intertwining operator and $\eta $-weak-pseudo-Hermiticity
generators [12]. A Hamiltonian $H$ is pseudo-Hermitian if it obeys the
similarity transformation $\eta \,H\,\eta ^{-1}=H^{\dagger },$ where $\eta $
is a Hermitian invertible linear operator and $(^{\dagger })$ denotes the
adjoint. The existence of real eigenvalues is realized to be associated\
with a non-Hermitian Hamiltonian provided that it is an $\eta $%
-pseudo-Hermitian:%
\begin{equation}
\eta \,H=H^{\dagger }\,\eta ,
\end{equation}%
with respect to the nontrivial "metric"operator $\eta =O^{\dagger }O$, for
some linear invertible operator $O:\mathcal{H}{\small \rightarrow }\mathcal{H%
}$ ($\mathcal{H}$ is the Hilbert space). However, under some rather mild
assumptions, we may even relax $H$ to be an $\eta $-weak-pseudo-Hermitian by
not restricting $\eta $ to be Hermitian (cf., e.g., Bagchi and Quesne [17]),
and linear and/or invertible (cf., e.g., Solombrino [18], Fityo [19], and
Mustafa and Mazharimousavi [12,20]).

Whilst in the non-Hermitian pseudo-Hermitian Hamiltonians neighborhood
[13-23], the non-Hermitian $\mathcal{PT}$-symmetric Hamiltonians (i.e., a
Bender's and Boettcher's [13] initiative on the so called nowadays $\mathcal{%
PT}$-symmetric quantum mechanics) are unavoidably in point. They form a
subclass of the non-Hermitian pseudo-Hermitian Hamiltonians \ (where $%
\mathcal{P}$ denotes parity and $\mathcal{T}$ mimics the time reversal).
Namely, if $\mathcal{PT}H\mathcal{PT}=H$ and if $\mathcal{PT}\Phi \left(
x\right) =\pm \Phi \left( x\right) $ the eigenvalues turn out to be real.
However, if the latter condition is not satisfied the eigenvalues appear in
complex-conjugate pairs (cf., e.g., Ahmed in [13]).

In this work, we consider (in section 2) a complexified von Roos Hamiltonian
(1) (i.e., $\tilde{V}\left( x\right) \longrightarrow \tilde{V}\left(
x\right) +iW\left( x\right) $) regardless of the nature of the ordering of
the ambiguity parameters as to being "good" or "to-be-discarded" ones. A
Hermitian first-order differential PDM-intertwining operator is used to
obtain the corresponding non-Hermitian $\eta $-weak-pseudo-Hermitian
PDM-Hamiltonian. The related \emph{reference}-\emph{target} non-Hermitian $%
\eta $-weak-pseudo-Hermitian Hamiltonians' map is also given in the same
section. Yet, in connection with the resulting effective \emph{reference}
potential, a \emph{"user-friendly"} form is suggested (in section 3) to
serve for exact-solvability of some non-Hermitian $\eta $%
-weak-pseudo-Hermitian PDM-Hamiltonians. Such a \emph{user-friendly} form
turns out to imply that there is always a set of isospectral \emph{target }$%
\eta $-weak-pseudo-Hermitian PDM-Hamiltonians associated with \emph{"one"}
exactly-solvable \emph{reference }$\eta $-weak-pseudo-Hermitian
PDM-Hamiltonian. We use (in the same section) two illustrative examples
(i.e., a complexified $PT$-symmetric Scarf-II and a periodic-type $PT$%
-symmetric Samsonov-Roy potentials) as \emph{reference} models and report
the corresponding sets of isospectral $\eta $-weak-pseudo-Hermitian \emph{%
target }-Hamiltonians. Section 4 is devoted for the concluding remarks.

\section{An $\protect\eta $-intertwiner and $\protect\eta $%
-weak-pseudo-Hermitian Hamiltonians' \emph{reference-target} map}

A complexification of the potential $\tilde{V}\left( x\right) $ in (1) may
be achieved by the transformation $\tilde{V}\left( x\right) \longrightarrow 
\tilde{V}\left( x\right) +iW\left( x\right) $, where $\tilde{V}\left(
x\right) ,W\left( x\right) \in 
%TCIMACRO{\U{211d} }%
%BeginExpansion
\mathbb{R}
%EndExpansion
$ and $%
%TCIMACRO{\U{211d} }%
%BeginExpansion
\mathbb{R}
%EndExpansion
\ni x\in \left( -\infty ,\infty \right) $. Hence, Hamiltonian (1) becomes
non-Hermitian and reads%
\begin{equation}
H=-\mu \left( x\right) ^{2}\partial _{x}^{2}-2\mu \left( x\right) \mu
^{\prime }\left( x\right) \partial _{x}+\tilde{V}\left( x\right) +iW\left(
x\right) ,
\end{equation}%
with $\mu \left( x\right) =\pm 1/\sqrt{M\left( x\right) }$. A Hermitian
first-order intertwining PDM-differential operator (cf., e.g., Mustafa and
Mazharimousavi [12] on the detailed origin of this PDM-operator) of the form%
\begin{equation}
\eta =-i\,\left[ \mu \left( x\right) \,\partial _{x}+\mu ^{\prime }\left(
x\right) /2\right] +F\left( x\right) ;\text{ \ \ \ }F\left( x\right) ,\mu
\left( x\right) \in 
%TCIMACRO{\U{211d} }%
%BeginExpansion
\mathbb{R}
%EndExpansion
\end{equation}%
would result, when used in (4),%
\begin{equation}
W\left( x\right) =-\mu \left( x\right) F^{\prime }\left( x\right) ,
\end{equation}%
\begin{equation}
\tilde{V}\left( x\right) =-F\left( x\right) ^{2}-\frac{1}{2}\mu \left(
x\right) \mu ^{\prime \prime }\left( x\right) -\frac{1}{4}\mu ^{\prime
}\left( x\right) ^{2}+\alpha _{\circ }.
\end{equation}%
where $\alpha _{\circ }\in $ $%
%TCIMACRO{\U{211d} }%
%BeginExpansion
\mathbb{R}
%EndExpansion
$ is an integration constant. One may then recast $V\left( x\right) $ as%
\begin{eqnarray}
V\left( x\right)  &=&\alpha _{\circ }-F\left( x\right) ^{2}+\left( \frac{1}{2%
}+\beta \right) \mu \left( x\right) \mu ^{\prime \prime }\left( x\right)  
\notag \\
&&+\left[ 4\alpha \left( \alpha +\beta +1\right) +\beta +\frac{3}{4}\right]
\mu ^{\prime }\left( x\right) ^{2}.
\end{eqnarray}%
One should, nevertheless, be reminded that an anti-Hermitian first-order
operator of the form $\eta =\mu \left( x\right) \,\partial _{x}+\mu ^{\prime
}\left( x\right) /2+iF\left( x\right) $ will exactly do the same job (cf.,
e.g., Mustafa and Mazharimousavi [12]). Moreover, as a result of this
intertwining process, a non-Hermitian $\eta $-weak-pseudo-Hermitian
Hamiltonian is obtained.

We may now consider our non-Hermitian $\eta $-weak-pseudo-Hermitian
Hamiltonian in (5), along with (7) and (8), in the one-dimensional Schr\"{o}%
dinger equation%
\begin{equation}
H_{x}\,\psi \left( x\right) =E\,\psi \left( x\right) 
\end{equation}%
and construct the so-called \emph{reference}-\emph{target} $\eta $%
-weak-pseudo-Hermitian Hamiltonians' map (equation (10) is the so-called 
\emph{target }Schr\"{o}dinger equation). A task that would be achieved by
the substitution 
\begin{equation}
\psi \left( x\right) \,=\varphi \left( q\left( x\right) \right) /\sqrt{\mu
\left( x\right) },
\end{equation}%
to imply, with the requirement 
\begin{equation}
q^{\prime }\left( x\right) =1/\mu \left( x\right) 
\end{equation}%
that removes the first-order derivative $\partial _{q}\varphi \left(
q\right) $, a so-called \emph{reference }Schr\"{o}dinger equation 
\begin{equation}
-\partial _{q}^{2}\varphi \left( q\left( x\right) \right) +\left[ \tilde{V}%
_{eff}\left( q\left( x\right) \right) -E\right] \varphi \left( q\left(
x\right) \right) =0,
\end{equation}%
where%
\begin{eqnarray}
\tilde{V}_{eff}\left( q\left( x\right) \right)  &=&\left( \beta +1\right)
\mu \left( x\right) \mu ^{\prime \prime }\left( x\right) +\left[ 4\alpha
\left( \alpha +\beta +1\right) +\beta +1\right] \mu ^{\prime }\left(
x\right) ^{2}  \notag \\
&&-F\left( x\right) ^{2}+\alpha _{\circ }-i\mu \left( x\right) F^{\prime
}\left( x\right) .
\end{eqnarray}%
It is evident that our $\eta $-weak-pseudo-Hermitian \emph{reference}%
-Hamiltonian%
\begin{equation}
H_{q}=-\partial _{q}^{2}+\tilde{V}_{eff}\left( q\right) ,
\end{equation}%
of (13), shares exactly the same spectrum of the $\eta $%
-weak-pseudo-Hermitian \emph{target}-Hamiltonian%
\begin{equation}
H_{x}=-\mu \left( x\right) ^{2}\partial _{x}^{2}-2\mu \left( x\right) \mu
^{\prime }\left( x\right) \partial _{x}-\frac{1}{4}\mu ^{\prime }\left(
x\right) ^{2}-\frac{1}{2}\mu \left( x\right) \mu ^{\prime \prime }\left(
x\right) +\tilde{V}_{eff}\left( x\right) ,
\end{equation}%
defined in (5), (7) and (8), where%
\begin{equation*}
\tilde{V}_{eff}\left( x\right) =\alpha _{\circ }-F\left( x\right) ^{2}-i\mu
\left( x\right) F^{\prime }\left( x\right) ,
\end{equation*}%
and $H_{q}$ and $H_{x}$ are isospectral. Nevertheless, one should keep in
mind that $H_{q}$ and $H_{x}$ may very well interchange their roles as to
being a \emph{reference} or a \emph{target} Hamiltonians. That is, it might
just happen that $H_{x}$ is exactly-solvable and in this case $H_{x}$
becomes a \emph{reference}-Hamiltonian and $H_{q}$ plays the role of being a 
\emph{target}-Hamiltonian.

\section{PDM-functions admitting isospectrality}

It is obvious that the effective \emph{reference} potential in (14) suggests
that the choice of%
\begin{equation}
\left( \beta +1\right) \mu \left( x\right) \mu ^{\prime \prime }\left(
x\right) +\left[ 4\alpha \left( \alpha +\beta +1\right) +\beta +1\right] \mu
^{\prime }\left( x\right) ^{2}=0,
\end{equation}%
would imply a "\emph{user-friendly}" effective \emph{reference} potential of
the form 
\begin{equation}
\tilde{V}_{eff}\left( q\right) =\alpha _{\circ }-F\left( q\right)
^{2}-iF^{\prime }\left( q\right) .
\end{equation}%
Hence%
\begin{equation*}
\mu ^{\prime }\left( x\right) \mu \left( x\right) ^{\delta }=const.
\end{equation*}%
and 
\begin{equation}
\mu \left( x\right) =\left[ C_{1}x+C_{2}\right] ^{1/\left( \delta +1\right)
}\,;\text{ }\delta =\left[ 4\alpha +1+\frac{4\alpha ^{2}}{\beta +1}\right] ,
\end{equation}%
where $C_{1}$ and $C_{2}$ are two constants and $C_{1},C_{2}\in 
%TCIMACRO{\U{211d} }%
%BeginExpansion
\mathbb{R}
%EndExpansion
$ . Nevertheless, one should notice that the Ben Daniel's and Duke's ($%
\alpha =\gamma =0,$ $\beta =-1$) ordering (although $\beta =-1$ is not
allowed by (19) but satisfies (17)) has already been discussed by Mustafa
and Mazharimousavi [12]. Hence, the Ben Daniel's and Duke's ordering shall
not be considered in the forthcoming studies. Moreover, under such mass
settings, we may report that; for Gora's and Williams' ($\beta =\gamma =0,$ $%
\alpha =-1$) and Li's and Kuhn's ($\beta =\gamma =-1/2,$ $\alpha =0$)
orderings $\delta _{GW}=\delta _{LK}=1$, for Zhu's and Kroemer's ($\alpha
=\gamma =-1/2,$ $\beta =0$) ordering $\delta _{ZK}=0$, and for Mustafa's and
Mazharimousavi's ($\alpha =\gamma =-1/4,$ $\beta =-1/2$) ordering $\delta
_{MM}=1/2$.

Moreover, it is evident that the position-dependent-mass $M\left( x\right) $%
\ under the current settings is strictly determined through (17) and
consequently through (19) to read%
\begin{equation}
M\left( x\right) =\mu \left( x\right) ^{-2}=\left[ C_{1}x+C_{2}\right]
^{-2/\left( \delta +1\right) }.
\end{equation}%
Hence, one may safely conclude that this PDM-form identifies a class of
isospectral position-dependent-mass functions satisfying the effective \emph{%
reference} potential\emph{\ }$\tilde{V}_{eff}\left( q\right) $ of (18), for
each form of the $\eta $-weak-pseudo-Hermiticity generator $F\left( q\right) 
$, and implies\medskip 
\begin{equation}
q\left( x\right) =\dint\nolimits^{x}\mu \left( y\right) ^{-1}dy=\left\{ 
\begin{tabular}{lc}
$\medskip \frac{\left( \delta +1\right) }{\delta C_{1}}\left[ C_{1}x+C_{2}%
\right] ^{\delta /\left( \delta +1\right) }$ & $\text{; for }\delta \neq 0$
\\ 
$\medskip \frac{1}{C_{1}}\ln \left( C_{1}x+C_{2}\right) $ & ; $\text{for }%
\delta =0$%
\end{tabular}%
\right. .
\end{equation}%
However, it should be noted that this case (i.e., $M\left( x\right) $ is
strictly determined) is unlike the one we have very recently considered in
[12], where Ben Daniel's and Duke's ordering (i.e., $\alpha =\gamma =0,$ $%
\beta =-1$) was used and the position-dependent-mass was left arbitrary
instead (but, of course, a positive-valued function). Yet, one should
clearly observe that the form of our $\tilde{V}_{eff}\left( q\right) $ in
(18) depends only on the choice of our $\eta $-weak-pseudo-Hermiticity
generator $F\left( q\right) $. It is advised that such a choice should be
oriented so that an exactly-solvable $\eta $-weak-pseudo-Hermitian \emph{%
reference} Hamiltonian is obtained. Consequently, a set of exactly-solvable
isospectral $\eta $-weak-pseudo-Hermitian \emph{target}-Hamiltonians of (16)
would result and depend only on the class of the strictly determined
position-dependent-mass functions in (20). Two illustrative examples are in
order.

\subsection{A complexified $PT$-symmetric Scarf-II model}

Let us recollect that an $\eta $-weak-pseudo-Hermiticity generator (cf.,
e.g., Mustafa and Mazharimousavi [12]) of the form%
\begin{equation}
F\left( q\right) =-V_{2}\func{sech}q\Longrightarrow F^{\prime }\left(
q\right) =V_{2}\func{sech}q\tanh q
\end{equation}%
would yield (with $\alpha _{\circ }=0$) a \emph{reference }effective
complexified $PT$-symmetric Scarf-II potential of the form%
\begin{equation}
\tilde{V}_{eff}\left( q\right) =-V_{2}^{2}\func{sech}^{2}q-iV_{2}\func{sech}%
q\tanh q\text{ };\text{ \ }%
%TCIMACRO{\U{211d} }%
%BeginExpansion
\mathbb{R}
%EndExpansion
\ni V_{2}\neq 0\text{.}
\end{equation}%
Which, in turn, would imply a \emph{target} effective potential of the form%
\begin{equation}
\tilde{V}_{eff}\left( x\right) =-4V_{2}^{2}\frac{\,f\left( x\right) ^{2}}{%
\left( f\left( x\right) ^{2}+1\right) ^{2}}\mp 2iV_{2}\frac{f\left( x\right)
\left( f\left( x\right) ^{2}-1\right) }{\left( f\left( x\right)
^{2}+1\right) ^{2}},
\end{equation}%
where $f\left( x\right) =\pm \exp \left[ q\left( x\right) \right] $, with $%
q\left( x\right) $ given in (21). In this case, the \emph{target} effective
potentials in (24) form a set of isospectral $\eta $-weak-pseudo-Hermitian
Hamiltonians 
\begin{eqnarray}
H_{x} &=&-\mu \left( x\right) ^{2}\partial _{x}^{2}-2\mu \left( x\right) \mu
^{\prime }\left( x\right) \partial _{x}-\frac{1}{4}\mu ^{\prime }\left(
x\right) ^{2}-\frac{1}{2}\mu \left( x\right) \mu ^{\prime \prime }\left(
x\right)  \\
&&-4V_{2}^{2}\frac{\,f\left( x\right) ^{2}}{\left( f\left( x\right)
^{2}+1\right) ^{2}}\mp 2iV_{2}\frac{f\left( x\right) \left( f\left( x\right)
^{2}-1\right) }{\left( f\left( x\right) ^{2}+1\right) ^{2}}.  \notag
\end{eqnarray}%
All of which share (with $\mu \left( x\right) $ as defined in (19)) the same
eigenvalues readily reported in [12,17] as 
\begin{equation}
E_{n}=-\left[ \left\vert V_{2}\right\vert -n-\frac{1}{2}\right] ^{2}\text{ };%
\text{ \ }n=0,1,2,\cdots ,n_{\max }<\left( \left\vert V_{2}\right\vert
-1/2\right) .
\end{equation}

\subsection{A periodic-type $PT$-symmetric Samsonov-Roy model}

We may also recycle our $\eta $-weak-pseudo-Hermiticity generator

\begin{equation}
F(q)=-\frac{4}{3\cos ^{2}q-4}-\frac{5}{4},
\end{equation}%
that implies (with $\alpha _{\circ }=0$) an effective periodic-type $PT$%
-symmetric Samsonov's and Roy's [12,14] \emph{reference} potential%
\begin{equation}
\tilde{V}_{eff}(q)=-\frac{6}{\left[ \cos q+2i\sin q\right] ^{2}}-\frac{25}{16%
}\text{ };\text{ \ }%
%TCIMACRO{\U{211d} }%
%BeginExpansion
\mathbb{R}
%EndExpansion
\ni q\in \left( -\pi ,\pi \right) .
\end{equation}%
This results, in effect, a \emph{target} effective potential of the form%
\begin{equation}
\tilde{V}_{eff}(x)=-\frac{6}{\left[ g\left( x\right) -2i\mu \left( x\right)
g^{\prime }\left( x\right) \right] ^{2}}-\frac{25}{16},
\end{equation}%
where \ $g\left( x\right) =\cos \left( q\left( x\right) \right) $, $\mu
\left( x\right) $ and $q\left( x\right) $ are as given in (19) and (21),
respectively. Hence, the set of $\eta $-weak-pseudo-Hermitian \emph{target}
Hamiltonians%
\begin{eqnarray}
H_{x} &=&-\mu \left( x\right) ^{2}\partial _{x}^{2}-2\mu \left( x\right) \mu
^{\prime }\left( x\right) \partial _{x}-\frac{1}{4}\mu ^{\prime }\left(
x\right) ^{2}-\frac{1}{2}\mu \left( x\right) \mu ^{\prime \prime }\left(
x\right) \\
&&-\frac{6}{\left[ g\left( x\right) -2i\mu \left( x\right) g^{\prime }\left(
x\right) \right] ^{2}}-\frac{25}{16}  \notag
\end{eqnarray}%
are isospectral and share the eigenvalues [12,14]%
\begin{equation}
E_{n}=\frac{n^{2}}{4}-\frac{25}{16}\text{ };\text{ \ }n=1,3,4,5,\cdots ,
\end{equation}%
with a missing $n=2$ state (the details of which can be found in Samsonov
and Roy [14]).

\section{Concluding remarks}

As long as $\eta $-weak-pseudo-Hermitian Hamiltonians are in point, their
solvability-nature/type (i.e., e.g., exact-, quasi-exact-,
conditionally-exact-, etc.) is still fresh and not yet adequately explored.
Amongst is the $\eta $-weak-pseudo-Hermitian von Roos PDM-Hamiltonian. In
this work, we tried to (at least) partially fill this gap and add a flavour
into such solvability territories of the $\eta $-weak-pseudo-Hermitian
Hamiltonians associated with position-dependent-mass settings.

In addition to mapping our $\eta $-weak-pseudo-Hermitian \emph{target}%
-Hamiltonians $H_{x}$ into $\eta $-weak-pseudo-Hermitian \emph{reference}%
-Hamiltonians $H_{q}$ (that share the same spectra for $H_{x}$ and is
advised to be exactly-solvable), we have suggested a \emph{"user-friendly"}
form (in $\tilde{V}_{eff}\left( q\right) $ of (18)) for the \emph{%
reference-target} $\eta $-weak-pseudo-Hermitian PDM-Hamiltonians' map. The
usage of which is exemplified through a non-Hermitian $\mathcal{PT}$
-symmetric Scarf II and a non-Hermitian $\mathcal{PT}$ -symmetric
Samsonov-Roy periodic-type models. It is observed that for each of these
models there is a set of exactly-solvable isospectral \emph{target} $\eta $%
-weak-pseudo-Hermitian PDM-Hamiltonians (documented in (25) for Scarf II and
in (30) for Samsonov-Roy). Hereby, it should be noted that the
isospectrality among the $\eta $-weak-pseudo-Hermitian Hamiltonians in (16)
is only manifested by the PDM choice of (17).

Of course there are other choices that might lead to some "user friendly"
forms of the effective potential in (14). The feasibility of the associated
isospectrality should always be explored, therefore. For example, the choice%
\begin{equation}
F\left( x\right) =\mu ^{\prime }\left( x\right) \Longrightarrow \mu \left(
x\right) =\dint\nolimits^{x}F\left( y\right) dy,
\end{equation}%
would imply an effective potential of the form%
\begin{equation}
\tilde{V}_{eff}\left( q\right) =-iF^{\prime }\left( q\right) +\left( \beta
+1\right) F^{\prime }\left( q\right) +\left[ 4\alpha \left( \alpha +\beta
+1\right) +\beta \right] F\left( q\right) ^{2}+\alpha _{\circ },
\end{equation}%
Apart from the ambiguity parameters' setting of $\beta =-1$ (and
consequently $\alpha =\gamma =0$ by the von Roos constraint $\alpha +\beta
+\gamma =-1$) considered by Mustafa and Mazharimousavi in [12], we were
unlucky to find any illustrative example that can be classified as
"successful" for such an effective potential form (33). Nonetheless, the
corresponding \emph{target} isospectral set of $\eta $-weak-pseudo-Hermitian
PDM-Hamiltonians is anticipated to be feasibly large (as documented by (32))
and not restricted to the position-dependent-mass form (unlike the case of $%
\tilde{V}_{eff}(q)$ of (18), which is restricted to the
position-dependent-mass function $M\left( x\right) $ in (20)).

Moreover, we may report that a generating function $F(q)=a\exp \left(
-q\right) $ would lead to (with $\alpha _{\circ }=0$) to%
\begin{equation}
\tilde{V}_{eff}\left( q\right) =-a^{2}\exp \left( -2q\right) +ia\exp \left(
-q\right)
\end{equation}%
of (18), and%
\begin{equation}
\tilde{V}_{eff}\left( q\right) =a^{2}\left[ 4\alpha \left( \alpha +\beta
+1\right) +\beta \right] \exp \left( -2q\right) -a\left( \beta +1-i\right)
\exp \left( -q\right)
\end{equation}%
of (33). The bound-states of the former in (32) (a non-Hermitian Morse
model) are reported to form an empty set of eigenvalues and, hence, labeled
as "unfortunate" for it leads to an empty set of isospectral $\eta $%
-weak-pseudo-Hermitian \emph{target}-Hamiltonians (cf., e.g., Mustafa and
Mazharimousavi [12], Bagchi and Quesne [22], and Ahmed [23]). The latter in
(33), on the other hand, does not fit into any of the "so-far-known"
exactly-solvable non-Hermitian Morse-type models, to the best of our
knowledge. These two models form open problems (if their bound-state
solutions exist at all), therefore.

Finally, one may add that the current strictly-determined set of \emph{target%
} effective potentials $\tilde{V}_{eff}\left( x\right) $ in (24) forms a
subset of the \emph{target} effective potentials reported in equations (25)
and (26) by Mustafa and Mazharimousavi [12]. Similar trend is also observed
for $\tilde{V}_{eff}\left( x\right) $ in (29) as it forms a subset of the
effective potentials in equations (34) and (35) of [12]. Hence, the scenario
of the \emph{energy-levels crossing} and the feasible manifestation of the 
\emph{flown away states} discussed in [12] remains effective, as long as the
our two illustrative examples are concerned.

\begin{acknowledgement}
We would like to thank the referees for their valuable comments and
suggestions.\newpage
\end{acknowledgement}


\begin{thebibliography}{99}
\bibitem{} Quesne C 2006 Ann. Phys. \textbf{321} 1221

Quesne C and Tkachuk V M \ 2004 J. \ Phys. \textbf{A}; Math and Gen \textbf{%
37} 4267

Tanaka T 2006 J. Phys. \textbf{A}; Math and Gen \textbf{39} 219

von Roos O 1983 Phys. Rev. \textbf{B 27} 7547

Gang C 2004 Phys Lett \textbf{A 329} 22

Jiang L, Yi L Z and Jia C S 2005 Phys. Lett. \textbf{A 345} 279

Mustafa O and Mazharimousavi S H 2006 Phys. Lett. \textbf{A 358} 259 (arXiv:
quant-ph/0603134)

\bibitem{} Mustafa O and Mazharimousavi S H 2006 J. Phys. \textbf{A}: Math.
Gen. \textbf{39} 10537 (arXiv: math-ph/0602044)

Alhaidari A D 2003 Int. J. Theor. Phys. \textbf{42} (2003) 2999

Alhaidari A D 2002 Phys. Rev. \textbf{A 66} 042116

\bibitem{} Puente A and Casas M 1994 Comput. Mater Sci. \textbf{2} 441

Mustafa O and Mazharimousavi S H 2007 Int. J. Theor. Phys. in press (arXiv:
quant-ph/0607158)

\bibitem{} Bastard G: \emph{"Wave Mechanics Applied to Semiconductor
Heterostructures" ,} (1988) Les Editions de Physique, Les Ulis

\bibitem{} Serra L I and Lipparini E 1997 Europhys. Lett. \textbf{40} 667

\bibitem{} Einevoll G. T. and Hemmer P. C. 1988 J. Phys. \textbf{C}: Solid
State Phys. \textbf{21} L1193.

Burt M. G. 1992 J. Phys. Condens. Matter \textbf{4} 6651.

Geller M. R. and Kohn W. 1993 Phys. Rev. Lett. \textbf{70} 3103.

Einevoll G. T. \ 1990 Phys. Rev. \textbf{B 42} 3497

Borges J S\'{a} et al 1988 Phys. Rev. \textbf{A 38 }3101

Ko\c{c} R., \c{S}ahino\^{g}lu G. and Koca M. 2005, Eur. Phys. J. \textbf{B} 
\textbf{48} 583 (arXive: quant-ph/0510172)

\bibitem{} de Souza Dutra A \ and Almeida C A S 2000 Phys Lett. \textbf{A 275%
} 25

\bibitem{} Gora T and Williams F 1969 Phys. Rev. \textbf{177} 1179.

\bibitem{} Ben Daniel D. J. and Duke C. B. 1966 Phys. Rev. \textbf{152} 683

\bibitem{} Zhu Q.-G. and Kroemer H. 1983 Phys. Rev. \textbf{B 27} 3519.

\bibitem{} Li T. and Kuhn K. J. 1993 Phys. Rev. \textbf{B 47} 12760

\bibitem{} Mustafa O and Mazharimousavi S H 2007 Int. J. Theor. Phys. in
press (arXiv: quant-ph/0607030)

\bibitem{} Bender C M and Boettcher S 1998 Phys. Rev. Lett. {\ }\textbf{80}
5243

Bender C M, Boettcher S and Meisinger P N 1999 J. Math. Phys. \textbf{40}
2201

Bagchi B, Cannata F and Quesne C 2000 Phys. Lett. \textbf{A 269} 79

Ahmed Z 2001 Phys. Lett. \textbf{A 282 }343

Ahmed Z 2001 Phys. Lett. \textbf{A 287 }295

Ahmed Z 2007 Phys. Lett. \textbf{A }in press.

\bibitem{} Khare A and Mandal B P 2000 Phys. Lett. \textbf{A 272 }53

Buslaev V and Grecchi V 1993 J. Phys.{\ A: Math. Gen. }\textbf{26} 5541

Znojil M and L\'{e}vai G 2000 Phys. Lett. A \textbf{271} 327

Bagchi B, Mallik S, Quesne C and Roychoudhury R 2001 Phys. Lett. \textbf{A
289} 34

Dorey P, Dunning C and Tateo R 2001 J. Phys. A: Math. Gen. \textbf{4} 5679

Kretschmer R and Szymanowski L 2004 Czech. J.Phys \textbf{54} 71

Znojil M, Gemperle F and Mustafa O 2002 J. Phys. A: Math. Gen. \textbf{35}
5781

Mustafa O and Znojil M 2002 J. Phys. A: Math. Gen. \textbf{35} 8929

Samsonov B F and Roy P 2005 J. Phys. \textbf{A}: Math. Gen. \textbf{38} L249.

\bibitem{} Mostafazadeh A 2002 J. Math. Phys. \textbf{43 }2814

Mostafazadeh A 2002 Nucl.Phys. \textbf{B 640} 419

Mostafazadeh A 2002 J. Math. Phys. \textbf{43 }205

Mostafazadeh A 2002 J. Math. Phys. \textbf{43} 3944

Mostafazadeh A 2003 J. Math. Phys. \textbf{44 }974

Mostafazadeh A 2005 J. Phys. \textbf{A}: Math. Gen. \textbf{38} 3213

\bibitem{} Sinha A and Roy P 2004 Czech. J. Phys. \textbf{54} 129

Jiang L, Yi L Z and Jia C S 2005 Phys Lett \textbf{A 345} 279

Mandal B P 2005 Mod. Phys. Lett. \textbf{A 20 }655

Znojil M, B\'{\i}la H and Jakubsky V 2004 Czech. J. Phys. \textbf{54} 1143

Mostafazadeh A and Batal A 2004 J. Phys.\textbf{A}: Math. Gen. \textbf{37 }%
11645

\bibitem{} Bagchi B and Quesne C 2002 Phys. Lett. \textbf{A 301} 173

\bibitem{} Solombrino L 2002 J. Math. Phys. \textbf{43} 5439

\bibitem{} Fityo T V 2002 J. Phys. A: Math. Gen. \textbf{35} 5893

\bibitem{} Mustafa O and Mazharimousavi S H 2006 Czech. J. Phys. \textbf{56 }%
967 (arXiv: quant-ph/0603237)

Mustafa O and Mazharimousavi S H 2006 Phys. Lett. \textbf{A 357} 295 (arXiv:
quant-ph/0604106)

\bibitem{} Mustafa O and Mazharimousavi S H 2006 (arXiv: hep-th/0601017)

\bibitem{} Bagchi B and Quesne C 2002 Phys. Lett. \textbf{A 301} 173

\bibitem{} Ahmed Z 2001 Phys. Lett. \textbf{A 290 }19
\end{thebibliography}
\end{document}